\documentclass[aps,preprint,tightenlines,groupedaddress,nofootinbib,byrevtex]{revtex4}

\usepackage{graphicx,comment}

\usepackage{amssymb,latexsym}
\usepackage{amsmath,amsbsy}

\begin{document}

\unitlength=1mm

\def\a{{\alpha}}
\def\b{{\beta}}
\def\d{{\delta}}
\def\D{{\Delta}}
\def\e{{\epsilon}}
\def\g{{\gamma}}
\def\G{{\Gamma}}
\def\k{{\kappa}}
\def\l{{\lambda}}
\def\L{{\Lambda}}
\def\m{{\mu}}
\def\n{{\nu}}
\def\o{{\omega}}
\def\O{{\Omega}}
\def\S{{\Sigma}}
\def\s{{\sigma}}
\def\th{{\theta}}

\def\ol#1{{\overline{#1}}}

\def\Dslash{D\hskip-0.65em /}
\def\qslash{D\hskip-0.65em /}

\def\CPT{{$\chi$PT}}
\def\QCPT{{Q$\chi$PT}}
\def\PQCPT{{PQ$\chi$PT}}
\def\tr{\text{tr}}
\def\str{\text{str}}
\def\diag{\text{diag}}
\def\order{{\mathcal O}}

\def\cC{{\mathcal C}}
\def\cB{{\mathcal B}}
\def\cT{{\mathcal T}}
\def\cQ{{\mathcal Q}}
\def\cL{{\mathcal L}}
\def\cO{{\mathcal O}}
\def\cA{{\mathcal A}}
\def\cH{{\mathcal H}}
\def\cR{{\mathcal R}}

\def\eqref#1{{(\ref{#1})}}

\preprint{NT@UW-03-021}

\title{Baryon Decuplet to Octet Electromagnetic 
       Transitions in \\
       Quenched and Partially Quenched Chiral Perturbation Theory}

\author{Daniel Arndt}
\email[]{arndt@phys.washington.edu}
\author{Brian C. Tiburzi}
\email[]{bctiburz@phys.washington.edu}
\affiliation{Department of Physics, Box 351560, 
             University of Washington, Seattle, WA 98195-1560, USA}

\date{\today}

\begin{abstract}
We calculate baryon decuplet to octet
electromagnetic transition form factors
in quenched and
partially quenched chiral perturbation theory.
We work in the isospin limit of $SU(3)$ flavor, 
up to next-to-leading order in the chiral expansion,
and to leading order in the heavy baryon expansion.
Our results are necessary for proper extrapolation
of lattice calculations of these transitions.
We also derive expressions for the case
of $SU(2)$ flavor
away from the isospin limit.
\end{abstract}

\pacs{}


\maketitle

\section{Introduction}
The study of the baryon
decuplet to octet electromagnetic transitions
provides important insight into the strongly interacting
regime of QCD.
Spin-parity selection rules for these
transitions allow for magnetic dipole (M1),
electric quadrupole (E2), and Coulumb quadrupole (C2) amplitudes.
Understanding these amplitudes, both in theory and experiment,
gives insight into the ground state wavefunctions
of the lowest lying baryons.
For example,
in the transition of the $\D(1232)$ to the nucleon,
if both baryon wavefunctions are spherically symmetric
then the E2 and C2 amplitudes vanish.
Experimentally, M1 is seen to be the dominant amplitude.
However, 
recent experimental measurements of the quadrupole amplitudes 
in the $\D\to N\g$ transition%
~\cite{Mertz:1999hp,Joo:2001tw}
show that the quadrupole amplitudes E2 and C2 are likely
non-zero.
This has revitalized the discussion as to the mechanism
for deformation of the baryons.
Although we expect more experimental data in the future,
progress 
will be slower for the remaining transitions as the
experimental difficulties are significant.

First-principle lattice QCD calculations of these matrix elements
can provide a theoretical explanation of these
experimental results.
In fact, 
the experimental difficulties may force us to rely
on lattice data 
for the non-nucleonic transitions.
Recently several such lattice calculations%
~\cite{Alexandrou:2002pw,Alexandrou:2003ea},
which improve upon
an earlier one~\cite{Leinweber:1993pv}, have appeared.
Although these calculations still largely employ 
the quenched approximation of 
QCD,
we expect partially quenched calculations
to be performed
in the near future.
Unfortunately 
now and foreseeably, these lattice calculations
cannot be performed with the physical masses of the light quarks
as the calculation time would be prohibitively long.
Therefore, to make physical predictions,
it is necessary 
to extrapolate from the heavier quark masses
used on the lattice 
(currently on the order of the strange quark mass) down to
the physical light quark masses.
Chiral perturbation theory (\CPT) provides model-independent
input for the behavior of observables as a function of quark masses.

For lattice calculations that use the quenched approximation
of QCD (QQCD), 
where the fermion determinant that
arises from the path integral is set equal to one,  
quenched chiral perturbation theory (\QCPT)%
~\cite{Morel:1987xk,Sharpe:1992ft,Bernard:1992ep,
Bernard:1992mk,Golterman:1994mk,Sharpe:1996qp,Labrenz:1996jy}
has been developed to aid in
the extrapolation.  
The problem with the quenched approximation is
that the Goldstone boson singlet,
the $\eta'$, which is heavy in QCD, 
remains light in QQCD and must be retained in \QCPT, 
requiring the addition of new operators and hence new
low-energy constants in the Lagrangian.
In general, the low-energy constants appearing in the \QCPT\
Lagrangian are unrelated to those in \CPT\ and
extrapolated quenched lattice data is unrelated to QCD.
In fact, several examples show that
the behavior of meson loops near the chiral limit is
frequently
misrepresented in \QCPT%
~\cite{Booth:1994rr,Kim:1998bz,Savage:2001dy,Arndt:2002ed,
Arndt:2003ww,Arndt:2003we}.

These problems of QQCD can be remedied by using partially quenched 
lattice QCD (PQQCD).  
Unlike QQCD, where the masses of quarks not connected to
external sources are set to infinity,
these ``sea quark'' masses are kept finite in PQQCD.
The masses of the sea quarks can be varied independently
of the valence quark masses;
usually they are chosen to be heavier. 
By keeping the sea quarks as dynamical degrees of freedom,
the fermion determinant is no longer equal to one
and needs to be computed. 
However, by efficaciously giving the sea quarks larger masses 
it is much less costly to calculate.
Moreover,
since PQQCD retains a $U(1)_A$ anomaly,
the equivalent to the singlet field in QCD is heavy (on the order
of
the chiral symmetry breaking scale $\L_\chi$) and can be integrated out%
~\cite{Sharpe:2000bn,Sharpe:2001fh}.
As a consequence, 
the low-energy constants appearing in 
partially quenched chiral perturbation theory (\PQCPT)%
~\cite{Bernard:1994sv,Sharpe:1997by,Golterman:1998st,Sharpe:1999kj,
Sharpe:2000bn,Sharpe:2000bc,Sharpe:2001fh,Shoresh:2001ha},
which is the low-energy effective theory of PQQCD,
are the same
as those appearing in \CPT.
By fitting \PQCPT\ to partially quenched
lattice data, one can determine these constants  
and actually make physical predictions for QCD.
\PQCPT\ has been used recently to study 
heavy meson~\cite{Savage:2001jw} and
octet baryon observables%
~\cite{Chen:2001yi,Beane:2002vq,Savage:2002fm,Leinweber:2002qb,Arndt:2003ww}.
The available lattice calculations for the 
$\D\to N\g$ transition%
~\cite{Alexandrou:2002pw,Alexandrou:2003ea}
use the quenched approximation;
there are currently no partially quenched simulations.
However, 
given the recent progress that lattice gauge theory has made
in the one-hadron sector and the prospect of
simulations  
in the two-hadron sector%
~\cite{LATTICEproposal1,LATTICEproposal2,
Beane:2002np,Beane:2002nu,Arndt:2003vx},
we expect to see partially quenched calculations of these
form factors in the near future.

This paper is organized as follows.
First, in Section~\ref{sec:PQCPT}, 
we briefly review \PQCPT\ including the treatment
of the baryon octet and decuplet in the heavy baryon approximation%
~\cite{Jenkins:1991jv,Jenkins:1991ne}.
Since we will use the conventions used in our
recent related work on the octet and decuplet baryons%
~\cite{Arndt:2003ww,Arndt:2003we}
we will keep this section brief.
In Section~\ref{sec:ff} we calculate
baryon decuplet to octet transition form factors
in both \QCPT\ and \PQCPT\
up to next-to-leading (NLO) order in the chiral expansion
and keep contributions to lowest order in the heavy baryon mass, $M_B$.
These calculations are done in the 
isospin limit of $SU(3)$ flavor.  
For completeness we also provide the
\PQCPT\ 
results for the transitions using the $SU(2)$ chiral
Lagrangian with non-degenerate quarks in the Appendix.
In Section~\ref{sec:conclusions} we conclude.

\section{\label{sec:PQCPT}\PQCPT}
In PQQCD the quark part of the Lagrangian is written as%
~\cite{Sharpe:2000bn,Sharpe:2001fh,Sharpe:2000bc,Sharpe:1999kj,
Golterman:1998st,Sharpe:1997by,Bernard:1994sv,Shoresh:2001ha}
\begin{equation}\label{eqn:LPQQCD}
  {\cal L}
  =
  \sum_{j,k=1}^9
  \bar{Q}_j(i\Dslash-m_Q)_{jk} Q_k
\end{equation}
that differs from the QCD $SU(3)$ flavor
Lagrangian by the inclusion of 
three bosonic ghost quarks, $\tilde{u}$, $\tilde{d}$, and $\tilde{s}$,
and three fermionic sea quarks, $j$, $l$, and $r$,
in addition to the fermionic light valence quarks $u$, $d$, and $s$.
These nine quarks are in the fundamental representation of
the graded group $SU(6|3)$%
~\cite{BahaBalantekin:1981kt,BahaBalantekin:1981qy,BahaBalantekin:1982bk}
and have been 
accommodated in the nine-component vector
\begin{equation}
  Q=(u,d,s,j,l,r,\tilde{u},\tilde{d},\tilde{s})
\end{equation}
that obeys the graded equal-time commutation relation
\begin{equation} \label{eqn:commutation}
  Q^\a_i({\bf x}){Q^\b_j}^\dagger({\bf y})
  -(-1)^{\eta_i \eta_j}{Q^\b_j}^\dagger({\bf y})Q^\a_i({\bf x})
  =
  \d^{\a\b}\d_{ij}\d^3({\bf x}-{\bf y})
,\end{equation}
where $\a$ and $\b$ are spin and $i$ and $j$ are flavor indices.
The graded equal-time commutation relations for two $Q$'s and two
$Q^\dagger$'s can be written analogously.
The grading factor 
\begin{equation}
   \eta_k
   = \left\{ 
       \begin{array}{cl}
         1 & \text{for } k=1,2,3,4,5,6 \\
         0 & \text{for } k=7,8,9
       \end{array}
     \right.
\end{equation}
takes into account the different statistics for
fermionic and bosonic quarks.
The quark mass matrix is given by 
\begin{equation}
  m_Q=\text{diag}(m_u,m_d,m_s,m_j,m_l,m_r,m_u,m_d,m_s)
\end{equation}
so that diagrams with closed ghost quark loops cancel 
those with valence quarks.
Effects of virtual quark loops are,
however, present due to the contribution of the finite-mass 
sea quarks. 

As has been recently realized~\cite{Golterman:2001yv}, 
the light quark electric charge matrix $\cQ$ is not uniquely
defined in PQQCD and the only
constraint one imposes is for
$\cQ$ to have vanishing
supertrace so that, as in QCD, no new operators
involving the singlet component are introduced.
Following~\cite{Chen:2001yi} we use
\begin{equation}
  \cQ
  =
  \diag
  \left(
    \frac{2}{3},-\frac{1}{3},-\frac{1}{3},q_j,q_l,q_r,q_j,q_l,q_r
  \right)
.\end{equation}
QCD is recovered in the limit 
$m_j\to m_u$, $m_l\to m_d$, and $m_r\to m_s$
independently of the $q$'s.

For massless quarks,
the Lagrangian in Eq.~(\ref{eqn:LPQQCD}) exhibits a graded symmetry
$SU(6|3)_L \otimes SU(6|3)_R \otimes U(1)_V$ that is assumed 
to be spontaneously broken down to $SU(6|3)_V \otimes U(1)_V$. 
The low-energy effective theory of PQQCD that emerges by 
expanding about the physical vacuum state is \PQCPT.
The dynamics of the emerging 80~pseudo-Goldstone mesons 
can be described at lowest 
order in the chiral expansion by the $\order(E^2)$ Lagrangian%
\footnote{
Here, $E\sim p$, $m_\pi$ where $p$ is an external momentum.
}
\begin{equation}\label{eqn:Lchi}
  {\cal L} =
  \frac{f^2}{8}
    \str\left(D^\mu\Sigma^\dagger D_\mu\Sigma\right)
    + \l\,\str\left(m_Q\Sigma+m_Q^\dagger\Sigma^\dagger\right)
    + \a\partial^\mu\Phi_0\partial_\mu\Phi_0
    - \mu_0^2\Phi_0^2
\end{equation}
where
\begin{equation} \label{eqn:Sigma}
  \Sigma=\exp\left(\frac{2i\Phi}{f}\right)
  = \xi^2,
\quad
  \Phi=
    \left(
      \begin{array}{cc}
        M & \chi^{\dagger} \\ 
        \chi & \tilde{M}
      \end{array}
    \right)
,\end{equation}
$f=132$~MeV,
and the gauge-covariant derivative is
$D_\mu\S=\partial_\mu\S+ie\cA_\mu[\cQ,\S]$.
The str() denotes a supertrace over flavor indices.
The $M$, $\tilde{M}$, and $\chi$ are matrices
of pseudo-Goldstone bosons with quantum numbers of $q\ol{q}$ pairs,
pseudo-Goldstone bosons with quantum numbers of 
$\tilde{q}\ol{\tilde{q}}$ pairs, 
and pseudo-Goldstone fermions with quantum numbers of $\tilde{q}\ol{q}$ pairs,
respectively.
$\Phi$ is defined in the quark basis and normalized such that
$\Phi_{12}=\pi^+$ (see, for example, \cite{Chen:2001yi}).
Upon expanding the Lagrangian in \eqref{eqn:Lchi} one finds that
to lowest order
the mesons with quark content $Q\bar{Q'}$
are canonically normalized when
their masses are given by
$m_{QQ'}^2=\frac{4\lambda}{f^2}(m_Q+m_{Q'})$.

The flavor singlet field given by $\Phi_0=\str(\Phi)/\sqrt{6}$
is, in contrast to the \QCPT\ case, rendered heavy by the $U(1)_A$
anomaly
and can therefore be integrated out in \CPT.
Analogously its mass $\mu_0$ can be taken to be 
on the order of the chiral symmetry breaking scale, 
$\mu_0\to\Lambda_\chi$.  
In this limit the 
flavor singlet propagator becomes independent of the
coupling $\a$ and 
deviates from a simple pole form~\cite{Sharpe:2000bn,Sharpe:2001fh}.

Just as there are mesons in PQQCD with quark content
$\ol{Q}_iQ_j$ that contain 
valence, sea, and ghost quarks, there are baryons 
with quark compositions $Q_iQ_jQ_k$ that
contain all three types of quarks.
To this end, one decomposes the irreducible representations
of $SU(6|3)_V$ into 
irreducible representations of 
$SU(3)_{\text{val}} \otimes SU(3)_{\text{sea}} \otimes SU(3)_{\text{ghost}}
 \otimes U(1)$.
The method to construct the octet baryons is to use the
interpolating field
\begin{equation}
  \cB_{ijk}^\g
  \sim
  \left(Q_i^{\a,a}Q_j^{\b,b}Q_k^{\g,c}-Q_i^{\a,a}Q_j^{\g,c}Q_k^{\b,b}\right)
  \e_{abc}(C\g_5)_{\a\b}
.\end{equation}
The spin-1/2 baryon octet $B_{ijk}=\cB_{ijk}$,
where the
indices $i$, $j$, and $k$ are restricted to $1$-$3$,
is contained as a $(\bf 8,\bf 1,\bf1)$ of
$SU(3)_{\text{val}} \otimes SU(3)_{\text{sea}} \otimes SU(3)_{\text{ghost}}$
in the $\bf 240$ representation.
The octet baryons, written in the familiar two-index notation
\begin{equation}
  B=
    \left(
      \begin{array}{ccc}
        \frac{1}{\sqrt{6}}\L+\frac{1}{\sqrt{2}}\S^0 & \S^+ & p \\ 
        \S^- & \frac{1}{\sqrt{6}}\L-\frac{1}{\sqrt{2}}\S^0 & n \\
        \Xi^- & \Xi^0 & -\frac{2}{\sqrt{6}}\L
      \end{array}
    \right)
,\end{equation}
are embedded in $B_{ijk}$ as~\cite{Labrenz:1996jy}
\begin{equation}
  B_{ijk}
  =
  \frac{1}{\sqrt{6}}
  \left(
    \e_{ijl}B_{kl}+\e_{ikl}B_{jl}
  \right)
.\end{equation}
The remaining 
baryon states needed for our calculation have at most 
one ghost or one sea quark
and have been constructed explicitly in~\cite{Chen:2001yi}.

Similarly, 
the familiar spin-3/2 decuplet baryons are embedded
in the $\bf 165$.  
Here,
one uses the interpolating field
\begin{equation} \label{eqn:Tstate}
  \cT_{ijk}^{\a,\mu}
  \sim
  \left(
    Q_i^{\a,a}Q_j^{\b,b}Q_k^{\g,c}
    +Q_i^{\b,b}Q_j^{\g,c}Q_k^{\a,a}
    +Q_i^{\g,c}Q_j^{\a,a}Q_k^{\b,b}
  \right)
  \e_{abc}
  \left(C\g^\mu\right)_{\b\g}
\end{equation}
that describes the $\bf 165$ dimensional representation of $SU(6|3)_V$. 
The decuplet baryons $T_{ijk}$
are then readily embedded in $\cT$ by construction:
$T_{ijk}=\cT_{ijk}$, where
the indices $i$, $j$, and $k$ are restricted to $1$--$3$.
They transform as a $(\bf 10, \bf 1, \bf1)$ under
$SU(3)_{\text{val}} \otimes SU(3)_{\text{sea}} \otimes SU(3)_{\text{ghost}}$.
Because of Eqs.~(\ref{eqn:commutation}) and \eqref{eqn:Tstate}, $T_{ijk}$ is
a totally symmetric tensor.  
Our normalization convention is such that $T_{111}=\D^{++}$.
For the spin-3/2 baryons consisting of two valence and one ghost quark
or two valence and one sea quark, we use the states constructed in%
~\cite{Chen:2001yi}.

At leading order in the heavy baryon expansion, the 
free Lagrangian for the $\cB_{ijk}$ and 
$\cT_{ijk}$ is given by~\cite{Labrenz:1996jy}
\begin{eqnarray} \label{eqn:L}
  {\mathcal L}
  &=&
  i\left(\ol\cB v\cdot{\mathcal D}\cB\right)
  +2\a_M\left(\ol\cB \cB{\mathcal M}_+\right)
  +2\b_M\left(\ol\cB {\mathcal M}_+\cB\right)
  +2\sigma_M\left(\ol\cB\cB\right)\str\left({\mathcal M}_+\right)
                              \nonumber \\
  &&-i\left(\ol\cT^\mu v\cdot{\mathcal D}\cT_\mu\right)
  +\D\left(\ol\cT^\mu\cT_\mu\right)
  +2\g_M\left(\ol\cT^\mu {\mathcal M}_+\cT_\mu\right)
  -2\ol\sigma_M\left(\ol\cT^\mu\cT_\mu\right)\str\left({\mathcal M}_+\right)
,\end{eqnarray}
where 
${\mathcal M}_+
  =\frac{1}{2}\left(\xi^\dagger m_Q \xi^\dagger+\xi m_Q \xi\right)$.
The brackets in (\ref{eqn:L}) are shorthands for field
bilinear invariants originally employed in~\cite{Labrenz:1996jy}.
The Lagrangian 
describing the relevant interactions of the $\cB_{ijk}$ 
and $\cT_{ijk}$ 
with the pseudo-Goldstone mesons is
\begin{equation} \label{eqn:Linteract}
  {\cal L} 
  =
  2\a\left(\ol{\cB}S^\mu \cB A_\mu\right)
  +
  2\b\left(\ol{\cB}S^\mu A_\mu \cB\right)
  +
  \sqrt{\frac{3}{2}}\cC
  \left[
    \left(\ol{\cT}^\nu A_\nu \cB\right)+\text{h.c.}
  \right]  
  +  
  2{\mathcal H}\left(\ol{\cT}^\nu S^\mu A_\mu \cT_\nu\right) 
\end{equation}
where the axial-vector and vector meson fields $A^\mu$ and $V^\mu$
are defined in analogy to those in QCD,
$A^\mu=\frac{i}{2}(\xi\partial^\mu\xi^\dagger-\xi^\dagger\partial^\mu\xi)$
and
$V^\mu=\frac{1}{2}(\xi\partial^\mu\xi^\dagger+\xi^\dagger\partial^\mu\xi)$.
The latter appears in Eq.~\eqref{eqn:L} in the
covariant derivatives of $\cB_{ijk}$ and $\cT_{ijk}$ 
that both have the form
\begin{equation}
  ({\mathcal D}^\mu \cB)_{ijk}
  =
  \partial^\mu \cB_{ijk}
  +(V^\mu)_{il}\cB_{ljk}
  +(-)^{\eta_i(\eta_j+\eta_m)}(V^\mu)_{jm}\cB_{imk}
  +(-)^{(\eta_i+\eta_j)(\eta_k+\eta_n)}(V^\mu)_{kn}\cB_{ijn}
.\end{equation}
By restricting the indices of $\cB_{ijk}$ to
$i,j,k=1,2,3$ one can relate the
constants $\a$ and $\b$ to  
$D$ and $F$ that are used for the $SU(3)_{\text{val}}$
analogs of these terms in QCD and finds 
\begin{equation}
  \a=\frac{2}{3}D+2F,\quad
  \b=-\frac{5}{3}D+F
,\end{equation}
while $\cC$ and $\cH$ are the constants of QCD.

\section{\label{sec:ff}Baryon Decuplet to Octet Transition}
The electromagnetic baryon decuplet to octet
transitions
have been investigated previously in \CPT%
~\cite{Butler:1993pn,Butler:1993ht,Napsuciale:1997ny,Gellas:1998wx}.
Very recently there also has been renewed interest in these transitions
in the large $N_c$ limit of QCD~\cite{Jenkins:2002rj}. 
Here we calculate these transitions in
\PQCPT\ and \QCPT.
While we have reviewed \PQCPT\ briefly in the last section 
and our recent papers~\cite{Arndt:2003ww,Arndt:2003we},
for \QCPT\ we refer the reader to the literature%
~\cite{Morel:1987xk,Sharpe:1992ft,Bernard:1992ep,
Bernard:1992mk,Golterman:1994mk,Sharpe:1996qp,Labrenz:1996jy}.

Using the heavy baryon formalism%
~\cite{Jenkins:1991jv,Jenkins:1991ne}, 
transition matrix elements of the electromagnetic current 
$J^\rho$ between a decuplet baryon with momentum
$p'$ and an octet baryon with momentum $p$
can be parametrized as
\begin{equation}
  \langle{\ol B}(p)|J^\rho|T(p')\rangle
  =
  {\ol u}(p)\cO^{\rho\mu}u_\mu(p')
,\end{equation}
where $u_\mu(p)$ is a Rarita-Schwinger spinor for an on-shell decuplet
baryon satisfying $v^\mu u_\mu(p)=0$ and $S^\mu u_\mu(p)=0$.
The tensor $\cO^{\rho\mu}$ can be parametrized 
in terms of three independent,
Lorentz invariant, dimensionless form factors%
~\cite{Jones:1973ky}
\begin{eqnarray}
  \mathcal{O}^{\rho \mu} 
  &=&
  \frac{G_1(q^2)}{M_B}
  \left(q\cdot S g^{\mu\rho} -q^\mu S^\rho\right)
  +
  \frac{G_2(q^2)}{(2M_B)^2}
  \left(q\cdot v g^{\mu\rho}-q^\mu v^\rho\right)S\cdot q
                          \nonumber \\
  &&+
  \frac{G_3(q^2)}{4M_B^2\D}
  \left(q^2 g^{\mu\rho}-q^\mu q^\rho\right)S\cdot q
,\end{eqnarray}
where the momentum of the outgoing photon is $q = p' - p$.
Here we have adopted the normalization of the $G_3(q^2)$
form factor used in~\cite{Gellas:1998wx}
so that the leading contributions to all three form factors 
are of order unity
in the power counting.

Linear combinations of the above form 
factors at $q^2=0$ make the magnetic dipole, electric quadrupole, 
and Coulombic quadrupole moments,
\begin{eqnarray}
  G_{M1}(0)&=&\left(\frac{2}{3}-\frac{\D}{6M_B}\right)G_1(0) 
             +\frac{\D}{12M_B}G_2(0),   \nonumber \\
  G_{E2}(0)&=&\frac{\D}{6M_B}G_1(0)+\frac{\D}{12M_B}G_2(0), \nonumber \\
  G_{C2}(0)&=&\left(\frac{1}{3}+\frac{\D}{6M_B}\right)G_1(0)
             +\left(\frac{1}{6}+\frac{\D}{6M_B}\right)G_2(0)
               +\frac{1}{6}G_3(0)   
.\end{eqnarray}

\subsubsection{\PQCPT}
Let us first consider the transition form factors 
in \PQCPT.
Here, the leading tree-level contributions to the transition moments 
come from the
dimension-5 and dimension-6 operators%
\footnote{We use
$F_{\mu\nu}=\partial_\mu A_\nu-\partial_\nu A_\mu$.}
\begin{equation} \label{eqn:Lbob}
  \cL 
  =
  \sqrt{\frac{3}{2}}\mu_T \frac{i e}{2 M_B} 
  \left( \ol \cB  S^\mu \cQ \cT^\nu \right)F_{\mu \nu} 
  +  
  \sqrt{\frac{3}{2}}\mathbb{Q}_T \frac{e}{\L_\chi^2} 
  \left( \ol \cB  S^{\{\mu} \cQ \cT^{\nu\}} \right)
  v^\a \partial_\mu F_{\nu \a}
\end{equation}
where
the action of ${}^{\{}\ldots{}^{\}}$ on Lorentz indices produces 
the symmetric traceless part of the tensor, 
{\it viz.}, 
$\mathcal{O}^{\{\mu \nu\}} 
 = \mathcal{O}^{\mu \nu} + \mathcal{O}^{\nu \mu} - 
\frac{1}{2} g^{\mu\nu} \mathcal{O}^{\alpha}{}_{\alpha}$.
Here the PQQCD low-energy constants $\mu_T$ and $\mathbb{Q}_T$
have the same numerical values as in QCD.

\begin{figure}
  \includegraphics[width=0.80\textwidth]{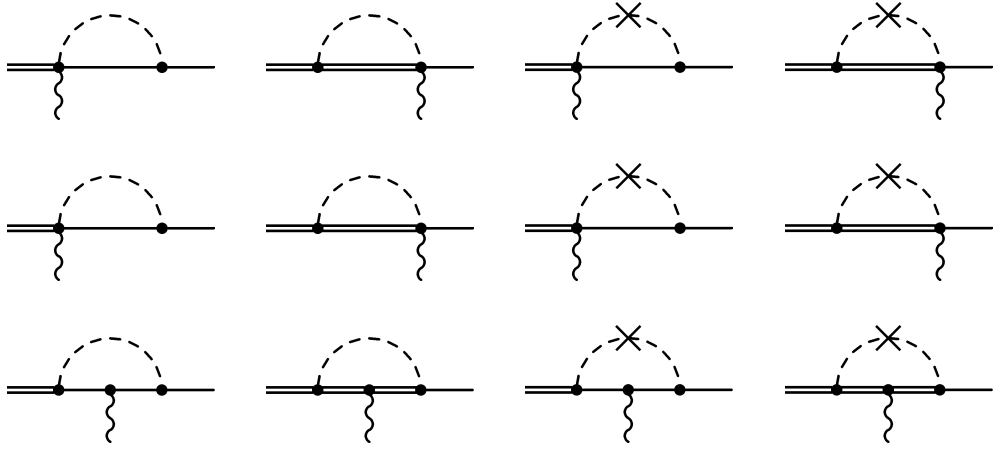}%
  \caption{\label{F:D2B-PQ-thatarezero}
     Loop diagrams that contribute to the transition
     moments but are zero to the order we are working.
     A thin (thick) solid line denotes an octet (decuplet)
     baryon whereas a dashed line denotes a meson.}
\end{figure}

\begin{figure}
  \includegraphics[width=0.40\textwidth]{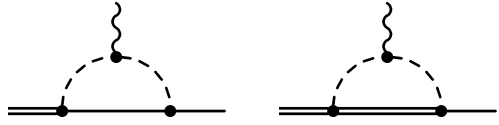}%
  \caption{\label{F:D2B-PQ}
     Loop diagrams contributing to the transition moments.}
\end{figure}

The NLO contributions in the chiral expansion
arise from the one-loop diagrams shown in
Figs.~(\ref{F:D2B-PQ-thatarezero}) and (\ref{F:D2B-PQ}).
However, because of the constraints satisfied by the on-shell
Rarita-Schwinger spinors, the diagrams in 
Fig.~(\ref{F:D2B-PQ-thatarezero}) are all identically zero.
Calculation of the diagrams in Fig.~(\ref{F:D2B-PQ}) gives
\begin{eqnarray}\label{eqn:G1}
  G_1(0)
  &=&
  \frac{\mu_T}{2}\a
  +
  \frac{M_B}{\L_\chi^2}4\cH\cC
  \sum_X\b_X^T
  \int_0^1 dx\,\left(1-\frac{x}{3}\right)
  \left[
    x\D\log\frac{m_X^2}{\mu^2}
      -m_X\cR\left(\frac{x\D}{m_X}\right)
  \right] \nonumber \\
  &&-
  \frac{M_B}{\L_\chi^2}4\cC(D-F)
  \sum_X\b_X^B
  \int_0^1 dx\,(1-x)
  \left[
    x\D\log\frac{m_X^2}{\mu^2}
      +m_X\cR\left(-\frac{x\D}{m_X}\right)
  \right]  
,\end{eqnarray}
\begin{eqnarray}
  G_2(0)
  &=&
  \frac{M_B^2}{\L_\chi^2}
  \Bigg\{
  -4\mathbb{Q}_T\a \nonumber \\
  &&\phantom{xxxxx}+
  16\cH\cC
  \sum_X\b_X^T
  \int_0^1 dx\,\frac{x(1-x)}{3}
  \left[
    \log\frac{m_X^2}{\mu^2}
      +\frac{x\D m_X}{m_X^2-x^2\D^2}
             \cR\left(\frac{x\D}{m_X}\right)
  \right]\nonumber \\
  &&-
  16\cC(D-F)
  \sum_X\b_X^B
  \int_0^1 dx\,x(1-x)
  \left[
    \log\frac{m_X^2}{\mu^2}
      -\frac{x\D m_X}{m_X^2-x^2\D^2}
           \cR\left(-\frac{x\D}{m_X}\right)
  \right]
  \Bigg\}  
,\end{eqnarray}
and
\begin{eqnarray}\label{eqn:G3}
  G_3(0)
  &=&
  -\frac{M_B^2}{\L_\chi^2}16
  \left[
  \cH\cC
  \sum_X\b_X^T
  \int_0^1 dx\,\frac{x(1-x)}{3}\left(x-\frac{1}{2}\right)
           \frac{\D m_X}{m_X^2-x^2\D^2}
             \cR\left(\frac{x\D}{m_X}\right)
  \right.\nonumber \\
  &&\left.
  +
  \cC(D-F)
  \sum_X\b_X^B
  \int_0^1 dx\,x(1-x)\left(x-\frac{1}{2}\right)
            \frac{\D m_X}{m_X^2-x^2\D^2}
           \cR\left(-\frac{x\D}{m_X}\right)
  \right]
,\end{eqnarray}
where
the function $\mathcal{R}(x)$ is given by
\begin{equation}
  \cR (x) 
  = 
  \sqrt{x^2 - 1} \, 
  \log 
  \frac{x - \sqrt{x^2 - 1 + i \epsilon}}{x + \sqrt{x^2 - 1 + i \epsilon}}
\end{equation}
and we have only kept loop contributions that
are non-analytic in the meson mass $m_X$.
The tree-level coefficients $\a$ are listed in Table~\ref{t:tree}
\begin{table} 
\caption{\label{t:tree}
Tree-level coefficients $\alpha$ in \CPT, \QCPT, and \PQCPT.}
\begin{tabular}{c | c  }
        \hline\hline
		              & $\alpha$ \\
	\hline
	$\D \to N \gamma$          
	& $\frac{1}{\sqrt{3}}$
	\\  
	$\Sigma^{*,+} \to \Sigma^+ \gamma$  
	& $-\frac{1}{\sqrt{3}}$
	\\  
	$\Sigma^{*,0} \to \Sigma^0 \gamma$            
	& $\frac{1}{2\sqrt{3}}$
	\\  
	$\Sigma^{*,0} \to \Lambda \gamma$            
	& $-\frac{1}{2}$	
	\\  
	$\Sigma^{*,-} \to \Sigma^- \gamma$            
	& $0$
	\\  
	$\Xi^{*,0} \to \Xi^{0} \gamma$            
	& $-\frac{1}{\sqrt{3}}$
	\\  
	$\Xi^{*,-} \to \Xi^- \gamma$            
	& $0$
        \\
        \hline\hline
\end{tabular}
\end{table} 
and the coefficients for the loop diagrams in Fig.~(\ref{F:D2B-PQ}),
$\b_X^T$ and $\b_X^B$,
are given in Tables~\ref{t:clebschT} and \ref{t:clebschB},
respectively.
In these tables we have listed values corresponding to the
loop meson with mass $m_X$.
As required, in the QCD limit the \PQCPT\ coefficients reduce
to those of \CPT.
\begin{turnpage}
\begin{table}
\caption{\label{t:clebschT}
The $SU(3)$ coefficients $\beta_X^T$ in \CPT\ and \PQCPT.}
\begin{ruledtabular}
\begin{tabular}{c | c c | c  c  c  c  c  c  c }
	& \multicolumn{2}{c |}{\CPT}  &  \multicolumn{7}{c}{\PQCPT} \\
	& $\pi$ & $K$  &   $\pi$   &   $K$   &  $\eta_s$  &   $ju$   
             &   $ru$   & $js$   & $rs$ 
	\\
	\hline
	$\D \to N \gamma$          
	&   $\frac{5}{3 \sqrt{3}}$    
	& $\frac{1}{3 \sqrt{3}}$   
	&   $\frac{1}{\sqrt{3}}$      
	&   $0$    
	&  $0$        
	&   $\frac{2}{3\sqrt{3}}$     
	&   $\frac{1}{3\sqrt{3}}$     
	&   $0$     
	& $0$ 
	\\
	$\Sigma^{*,+} \to \Sigma^+ \gamma$  
	&   $-\frac{1}{3 \sqrt{3}}$    
	& $-\frac{5}{3 \sqrt{3}}$   
	&   $\frac{1 - 3 q_{jl}}{9\sqrt{3}} $   
	&   $-\frac{11 - 3 q_{jl} + 3 q_r}{9\sqrt{3}}$    
	&  $\frac{1 + 3 q_r}{9\sqrt{3}}$        
	&   $-\frac{4 - 3 q_{jl}}{9\sqrt{3}}$     
	&   $-\frac{2 - 3 q_r}{9\sqrt{3}}$     
	&   $-\frac{2 + 3 q_{jl}}{9\sqrt{3}}$     
	& $-\frac{1 + 3 q_r}{9\sqrt{3}}$ 
	\\
 	$\Sigma^{*,0} \to \Sigma^0 \gamma$            
	&   $0$    
	& $\frac{1}{\sqrt{3}}$   
	&   $-\frac{1 - 3 q_{jl}}{9\sqrt{3}}$    
	&   $\frac{13 - 6 q_{jl} + 6 q_r}{18 \sqrt{3}}$    
	&  $- \frac{1 + 3 q_r}{9 \sqrt{3}}$   
    	&   $\frac{1 - 3 q_{jl}}{9\sqrt{3}}$     
	&   $\frac{1 - 6 q_r}{18 \sqrt{3}}$     
	&   $\frac{2 + 3 q_{jl}}{9\sqrt{3}}$     
	& $\frac{1 + 3 q_{r}}{9\sqrt{3}}$ 
	\\
	$\Sigma^{*,0} \to \Lambda \gamma$            
	&   $-\frac{2}{3}$    
	& $-\frac{1}{3}$   
	&   $-\frac{1}{3}$      
	&   $-\frac{1}{6}$    
	&  $0$        
	&   $-\frac{1}{3}$     
	&   $-\frac{1}{6}$     
	&   $0$     
	& $0$ 	
	\\
	$\Sigma^{*,-} \to \Sigma^- \gamma$            
	&   $-\frac{1}{3 \sqrt{3}}$    
	& $\frac{1}{3 \sqrt{3}}$   
	&   $-\frac{1 - 3 q_{jl}}{9\sqrt{3}}$      
	&   $\frac{2 - 3 q_{jl} + 3 q_r}{9 \sqrt{3}}$    
	&  $-\frac{1 + 3 q_{r}}{9\sqrt{3}}$   
     	&   $-\frac{2 + 3 q_{jl}}{9 \sqrt{3}}$     
	&   $-\frac{1 + 3 q_{r}}{9 \sqrt{3}}$     
	&   $\frac{2 + 3 q_{jl}}{9 \sqrt{3}}$     
	&    $\frac{1 + 3 q_{r}}{9 \sqrt{3}}$ 
	\\
	$\Xi^{*,0} \to \Xi^{0} \gamma$            
	&   $-\frac{1}{3 \sqrt{3}}$    
	& $-\frac{5}{3 \sqrt{3}}$   
	&   $\frac{1 - 3 q_{jl}}{9\sqrt{3}}$      
	&   $-\frac{11 - 3 q_{jl} + 3 q_r}{9 \sqrt{3}}$    
	&  $\frac{1 + 3 q_{r}}{9\sqrt{3}}$    
	&   $-\frac{4 - 3 q_{jl}}{9\sqrt{3}}$     
	&   $-\frac{2 - 3 q_{r}}{9\sqrt{3}}$     
	&   $-\frac{2 + 3 q_{jl}}{9\sqrt{3}}$     
	& $-\frac{1 + 3 q_{r}}{9\sqrt{3}}$ 
	\\
	$\Xi^{*,-} \to \Xi^- \gamma$            
	&   $-\frac{1}{3 \sqrt{3}}$    
	& $\frac{1}{3 \sqrt{3}}$   
	&   $-\frac{1 - 3 q_{jl}}{9\sqrt{3}}$      
	&   $\frac{2 - 3 q_{jl} + 3 q_r}{9 \sqrt{3}}$    
	&  $-\frac{1 + 3 q_{r}}{9\sqrt{3}}$     
	&   $-\frac{2 + 3 q_{jl}}{9\sqrt{3}}$     
	&   $-\frac{1 + 3 q_{r}}{9\sqrt{3}}$     
	&   $\frac{2 + 3 q_{jl}}{9\sqrt{3}}$     
	& $\frac{1 + 3 q_{r}}{9\sqrt{3}}$
\end{tabular}
\end{ruledtabular}
\end{table} 
\begin{table}
\caption{\label{t:clebschB}
The $SU(3)$ coefficients $\beta_X^B$ in \CPT\ and \PQCPT.}
\begin{ruledtabular}
\begin{tabular}{c | c c | c  c  c  c  c  c  c }
	& \multicolumn{2}{c |}{\CPT}  &  \multicolumn{7}{c}{\PQCPT} \\
	& $\pi$ & $K$  &   $\pi$   &   $K$   &  $\eta_s$  &   $ju$   
                  &   $ru$   & $js$   & $rs$ \\
	\hline
	$\D \to N \gamma$          
	&   $-\frac{D + F}{\sqrt{3}(D-F)} $    
	&   $- \frac{1}{\sqrt{3}} $   
	&   $\frac{D - 3 F}{\sqrt{3}(D-F)}$    
	&   $0$    
	&  $0$        
	&   $-\frac{2}{\sqrt{3}}$    
	&   $- \frac{1}{\sqrt{3}}$     
	&   $0$    
	&   $0$ 
	\\
	$\Sigma^{*,+} \to \Sigma^+ \gamma$  
	&   $\frac{1}{\sqrt{3}} $    
	&   $\frac{D + F}{\sqrt{3}(D-F)} $   
	&   $- \frac{1 - 3 q_{jl}}{3\sqrt{3}} $    
	&   $- \frac{D - 7 F}{3\sqrt{3}(D-F)} + \frac{q_{jl}-q_r}{\sqrt{3}}$  
	&  $- \frac{1 + 3 q_r}{3 \sqrt{3}}$        
	&   $\frac{4 - 3 q_{jl}}{3 \sqrt{3}}$    
	&   $\frac{2 - 3 q_r}{3 \sqrt{3}}$     
	&   $\frac{2 + 3 q_{jl}}{3\sqrt{3}}$    
	&   $\frac{1 + 3 q_r}{3\sqrt{3}}$ 
	\\
	$\Sigma^{*,0} \to \Sigma^0 \gamma$            
	&   $0$    
	&   $-\frac{D}{\sqrt{3}(D-F)} $   
	&   $\frac{1 - 3 q_{jl}}{3\sqrt{3}}$    
	&   $-\frac{D + 5 F}{6\sqrt{3}(D-F)} - \frac{q_{jl}-q_r}{\sqrt{3}}$  
	&   $\frac{1 + 3 q_r}{3 \sqrt{3}}$        
	&   $-\frac{1 - 3 q_{jl}}{3\sqrt{3}}$    
	&   $-\frac{1 - 6 q_{r}}{6\sqrt{3}}$     
	&   $-\frac{2+ 3 q_{jl}}{3\sqrt{3}}$    
	&   $-\frac{1+ 3 q_{r}}{3\sqrt{3}}$ 
	\\
	$\Sigma^{*,0} \to \Lambda \gamma$            
	&   $\frac{2D}{3(D-F)}$    
	&   $\frac{D}{3(D-F)}$   
	&   $-\frac{D-3F}{3(D - F )}$    
	&   $-\frac{D-3F}{6(D - F )}$    
	&  $0$        
	&   $1$    
	&   $\frac{1}{2}$     
	&   $0$    
	&   $0$ 
	\\
	$\Sigma^{*,-} \to \Sigma^- \gamma$            
	&   $\frac{1}{\sqrt{3}} $    
	&   $-\frac{1}{\sqrt{3}}$   
	&   $\frac{1 - 3 q_{jl}}{3\sqrt{3}}$    
	&   $-\frac{2 - 3 q_{jl} + 3 q_r}{3\sqrt{3}}$    
	&  $\frac{1+ 3 q_{r}}{3\sqrt{3}}$        
	&   $\frac{2 + 3 q_{jl}}{3\sqrt{3}}$    
	&   $\frac{1+ 3 q_{r}}{3\sqrt{3}}$     
	&   $-\frac{2 + 3 q_{jl}}{3\sqrt{3}}$    
	&   $-\frac{1+ 3 q_{r}}{3\sqrt{3}}$ 
	\\
	$\Xi^{*,0} \to \Xi^{0} \gamma$            
	&   $\frac{1}{\sqrt{3}} $    
	&   $\frac{D + F}{\sqrt{3}(D-F)} $   
	&   $-\frac{1 - 3 q_{jl}}{3\sqrt{3}}$    
	&   $- \frac{D - 7 F}{3\sqrt{3}(D-F)} + \frac{q_{jl}-q_r}{\sqrt{3}}$   
	&  $-\frac{1+ 3 q_{r}}{3\sqrt{3}}$        
	&   $\frac{4 - 3 q_{jl}}{3\sqrt{3}}$    
	&   $\frac{2- 3 q_{r}}{3\sqrt{3}}$     
	&   $\frac{2 + 3 q_{jl}}{3\sqrt{3}}$    
	&   $\frac{1 + 3 q_{r}}{3\sqrt{3}}$ 
	\\
	$\Xi^{*,-} \to \Xi^- \gamma$            
	&   $\frac{1}{\sqrt{3}} $    
	&   $-\frac{1}{\sqrt{3}} $   
	&   $\frac{1 - 3 q_{jl}}{3\sqrt{3}}$    
	&   $-\frac{2 - 3 q_{jl} + 3 q_r}{3\sqrt{3}}$    
	&  $\frac{1 + 3 q_{r}}{3\sqrt{3}}$        
	&   $\frac{2 + 3 q_{jl}}{3\sqrt{3}}$    
	&   $\frac{1 + 3 q_{r}}{3\sqrt{3}}$     
	&   $- \frac{2 + 3 q_{jl}}{3\sqrt{3}}$    
	&   $-\frac{1 + 3 q_{r}}{3\sqrt{3}}$ 
\end{tabular}
\end{ruledtabular}
\end{table} 
\end{turnpage}
It is comforting that the one-loop results for the
$G_3(q^2)$ form factor are finite. 
This is consistent with
the fact that one cannot write down a dimension-7
operator that contributes at the same order in the
chiral expansion as our one-loop result for $G_3(q^2)$.
The full one-loop $q^2$ dependence of these form factors
can easily be recovered by replacing
\begin{equation}
  m_X\to\sqrt{m_X^2-x(1-x)q^2}
.\end{equation}

Notice that the tree-level transitions 
$\S^{*,-} \to \S^- \gamma$ and 
$\Xi^{*,-} \to \Xi^- \gamma$
are zero because they are forbidden by 
$d\leftrightarrow s$
$U$-spin symmetry~\cite{Lipkin:1973rw}.
There is also symmetry between the
$\S^{*,+} \to \S^+ \gamma$ and 
$\Xi^{*,0} \to \Xi^0 \gamma$ transitions
as well as the
$\S^{*,-} \to \S^- \gamma$ and 
$\Xi^{*,-} \to \Xi^- \gamma$ transitions
that holds
to NLO in \CPT\ and \PQCPT.

\subsubsection{\QCPT}
The calculation of the transition moments can 
be repeated in \QCPT.  
At tree level, the operators in Eq.~\eqref{eqn:Lbob}
contribute, but their low-energy coefficients
cannot be matched onto QCD.  Therefore we annotate them with a ``Q''.
At the next order in the chiral expansion,
there are again contributions from
the loop diagrams in Fig.~(\ref{F:D2B-PQ}). 
The results are the same as in the partially quenched
theory, Eqs.\ \eqref{eqn:G1}--\eqref{eqn:G3}, with the
coefficients $\b_X^T$ and $\b_X^B$ replaced by
$\b_X^{T,Q}$ and $\b_X^{B,Q}/(D^Q-F^Q)$, 
which are listed in Table~\ref{t:clebschQ}.
\begin{table}
\caption{\label{t:clebschQ}
The $SU(3)$ coefficients $\beta_X^{B,Q}$ and $\beta_X^{T,Q}$ in \QCPT.}
\begin{tabular}{c | c c | c c }
	    & \multicolumn{2}{c |}{$\beta_X^{T,Q}$} & \multicolumn{2}{c}{$\beta_X^{B,Q}$}  \\ 
	              & $\pi$ & $K$  & $\pi$ & $K$ \\ 
	\hline
	$\D \to N \gamma$          
	&   $\frac{1}{\sqrt{3}}$    & $0$   
        & $\frac{1}{\sqrt{3}}(D^Q - 3 F^Q)$ & $0$
	\\  
	$\Sigma^{*,+} \to \Sigma^+ \gamma$  
	&   $0$    & $-\frac{1}{\sqrt{3}}$ 
 	& $0$ & $-\frac{1}{\sqrt{3}} (D^Q - 3 F^Q)$  
	\\  
	$\Sigma^{*,0} \to \Sigma^0 \gamma$            
	&   $0$    & $\frac{1}{2\sqrt{3}}$   
	& $0$ & $\frac{1}{2 \sqrt{3}}(D^Q - 3 F^Q)$
	\\  
	$\Sigma^{*,0} \to \Lambda \gamma$            
	&   $-\frac{1}{3}$    & $-\frac{1}{6}$   
	& $-\frac{1}{3} ( D^Q - 3 F^Q)$ & $-\frac{1}{6} (D^Q - 3 F^Q)$
	\\  
	$\Sigma^{*,-} \to \Sigma^- \gamma$            
	&   $0$    & $0$   
	&   $0$    & $0$
	\\  
	$\Xi^{*,0} \to \Xi^{0} \gamma$            
	&   $0$    & $-\frac{1}{\sqrt{3}}$   
	&   $0$ & $- \frac{1}{\sqrt{3}}(D^Q - 3 F^Q)$
	\\  
	$\Xi^{*,-} \to \Xi^- \gamma$            
	&   $0$    & $0$   
	& $0 $ & $0$
\end{tabular}
\end{table} 

\begin{figure}
  \includegraphics[width=0.40\textwidth]{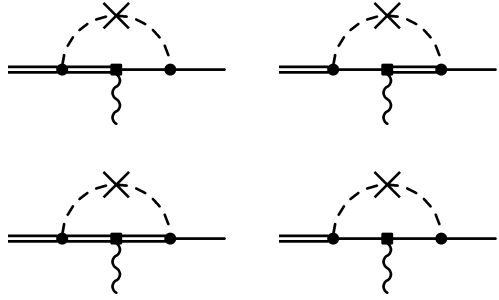}%
  \caption{\label{F:D2B-Q}
     Loop diagrams contributing to the transition form factors
     in \QCPT. The four diagrams correspond to terms involving the
     parameters $A_{XX'}$, $B_{XX'}$, $C_{XX'}$, and $D_{XX'}$
     in Eqs.\ \eqref{eqn:G1HP} and \eqref{eqn:G2HP}.}
\end{figure}

In addition,
there are contributions of the form $\mu_0^2\log m_q$
at the same order in the chiral expansion
that are artifacts of quenching.
These come from hairpin
wavefunction renormalization diagrams 
and from
the four loop diagrams in Fig.~(\ref{F:D2B-Q}).
In these diagrams the photon can couple to the 
baryon line via
\begin{eqnarray}\label{eqn:LDF}
  {\cal L}
  &=&
  \frac{ie}{2M_B}
  \left[
    \mu_\a^Q\left(\ol{\cB}[S_\mu,S_\nu]\cB\cQ\right)
    +\mu_\b^Q\left(\ol{\cB}[S_\mu,S_\nu]\cQ\cB\right)
  \right]
  F^{\mu\nu} \nonumber \\
  &&+
  \mu_c^Q \frac{ 3 i e }{M_B}  \big(\ol\cT_\mu \cQ \cT_\nu \big) F^{\mu \nu}
  - 
  \mathbb{Q}_{\text{c}}^Q \frac{3 e}{\L_\chi^2} 
  \big(\ol \cT{}^{\{\mu} \cQ \cT^{\nu\}} \big)  
  v^\alpha \partial_\mu F_{\nu \alpha}
\end{eqnarray}
and via the terms in Eq.~\eqref{eqn:Lbob} including
their hermitian conjugates (with quenched coefficients).%
\footnote{
Note that possible contributions from diagrams involving
\begin{equation}
  {\cal L}
  =
  \frac{e}{\L_\chi^2}
  \left[
    c_\a^Q(\ol{\cB}\cB\cQ)+c_\b^Q(\ol{\cB}\cQ\cB)
  \right]
  v_\mu\partial_\nu F^{\mu\nu}
  +
  c_c^Q \frac{3 e}{\L_\chi^2} 
  \big( \ol \cT{}^\sigma \cQ \cT_{\sigma} \big) 
  v_\mu \partial_\nu F^{\mu \nu}
\end{equation}
are identically zero.}
It is easier to work with the combinations $\mu_D^Q$ and $\mu_F^Q$ 
defined by
\begin{equation}
  \mu_\a^Q 
  = 
  \frac{2}{3} \mu_D^Q + 2 \mu_F^Q \quad \text{and} \quad 
  \mu_\b^Q 
  = 
  -\frac{5}{3} \mu_D^Q + \mu_F^Q
.\end{equation}
Although the argument presented in~\cite{Chow:1998xc}  
does not apply to the case of different initial
and final states, the axial hairpin
interactions still do not contribute 
simply because their presence requires
closed quark loops.
The hairpin
wavefunction renormalization diagrams have been
calculated in \QCPT\ for the 
baryon octet~\cite{Savage:2001dy} ($Z^Q_B$)
and decuplet~\cite{Arndt:2003we} ($Z^Q_T$) and
we do not reproduce them here.
We find
the hairpin contributions to the transition form factors
to be
\begin{eqnarray} \label{eqn:G1HP}
  G^{HP}_1(q^2)
  &=&
  \frac{\mu_T^Q}{2}\a\frac{Z_B^Q-1}{2}\frac{Z_T^Q-1}{2} \nonumber \\
  &&+
  \frac{\mu_0^2}{16\pi^2f^2}
  \sum_{X,X'}
  \Bigg[
    \frac{5}{108}\cH^Q\mu_T^Q A_{XX'}I_{XX'}
    -\frac{1}{18}\left(\cC^Q\right)^2\mu_T^QB_{XX'}I_{XX'}^{-\D,\D}
                    \nonumber \\
    &&\phantom{xxxx}
    -\frac{20}{27}\cH^Q\cC^QQ_T\mu_c^QC_{XX'}I_{XX'}^\D
    -\frac{2}{3}\cC^Q\left(Q_T\mu_F^Q+\a_D\mu_D^Q\right)D_{XX'}I^{\D}_{XX'}
  \Bigg]
,\end{eqnarray}
\begin{eqnarray} \label{eqn:G2HP}
  G^{HP}_2(q^2)
  &=&
  -4{\mathbb Q}_T^Q\a\frac{M_B^2}{\L_\chi^2}
   \frac{Z_B^Q-1}{2}\frac{Z_T^Q-1}{2}
                          \nonumber \\
  &&+
  \frac{\mu_0^2}{16\pi^2f^2}\frac{M_B^2}{\L_\chi^2}
  \sum_{XX'}
  \Bigg[
    \frac{2}{9}\cH^Q{\mathbb Q}_T^QA_{XX'}I_{XX'}
    +\frac{4}{3}\left(C^Q\right)^2
                      {\mathbb Q}_T^QB_{XX'}I^{-\D\,\D}_{XX'}
                         \nonumber \\
    &&\phantom{xxxxxxxxxxxxxxxx}
    -\frac{16}{9}\cH^Q\cC^QQ_T{\mathbb Q}_C^QC_{XX'}I^\D_{XX'}
  \Bigg]
,\end{eqnarray}
and $G^{HP}_3(q^2)=0$.
Thus in \QCPT:
$G_j^Q(q^2)=G_j^{PQ}(q^2)+G_j^{HP}(q^2)$,
where the $\b_X^T$ and $\b_X^B$ coefficients of
$G_j^{PQ}(q^2)$, Eqs.\ (\ref{eqn:G1})--(\ref{eqn:G3}),
are understood to be replaced by their quenched values 
$\b_X^{T,Q}$ and $\b_X^{B,Q}/(D^Q-F^Q)$.
Above we have used the shorthand notation
$I_{\eta_q\eta_{q^\prime}}=I(m_{\eta_q},m_{\eta_{q^\prime}},0,0,\mu)$,
$I^{\D}_{\eta_q\eta_{q^\prime}}=I(m_{\eta_q},m_{\eta_{q^\prime}},\D,0,\mu)$,
and
$I^{\D_1,\D_2}_{\eta_q\eta_{q^\prime}}
=I(m_{\eta_q},m_{\eta_{q^\prime}},\D_1,\D_2,\mu)$
for the function $I(m_1,m_2,\D_1,\D_2,\mu)$ that is given by
\begin{equation}
  I(m_1,m_2,\D_1,\D_2,\mu) 
  = 
  \frac{Y(m_1,\D_1,\mu) +Y(m_2,\D_2,\mu) - Y(m_1,\D_2,\mu) - Y(m_2,\D_1,\mu)}
       {(m_1^2 - m_2^2)(\D_1 - \D_2)}
\end{equation}
with
\begin{equation}
  Y(m,\D,\mu) 
  = 
  \D\left(m^2-\frac{2}{3}\D^2\right)\log\frac{m^2}{\mu^2} 
  +\frac{2}{3}m(\D^2-m^2)\mathcal{R}\left(\frac{\D}{m}\right)
.\end{equation}
The coefficients $A_{XX'}$, $B_{XX'}$, $C_{XX'}$, and $D_{XX'}$
are listed in Tables \ref{t:QclebschAB} and \ref{t:QclebschCD}.
\begin{table}
\caption{\label{t:QclebschAB}
The $SU(3)$  coefficients $A_{XX'}$ and $B_{XX'}$ in \QCPT.}
\begin{ruledtabular}
\begin{tabular}{c | c  c  c | c  c  c }
	& \multicolumn{3}{c |}{$A_{XX'}$}& \multicolumn{3}{c}{$B_{XX'}$}\\
        & $\eta_u \eta_u$ & $\eta_u \eta_s$  &   $\eta_s \eta_s$  & 
            $\eta_u \eta_u$ & $\eta_u \eta_s$  &   $\eta_s \eta_s$     \\
	\hline
	$\D \to N \gamma$          
	& $2\sqrt{3}(D^Q-3F^Q)$
	& $0$
	& $0$
	& $0$
	& $0$	
	& $0$
	\\  
	$\Sigma^{*,+} \to \Sigma^+ \gamma$  
	& $\frac{8}{\sqrt{3}}  F^Q$
	& $-\frac{4}{\sqrt{3}}  (  D^Q - 2 F^Q)$
	& $-\frac{2}{\sqrt{3}}  (D^Q - F^Q)$
	& $\frac{1}{3 \sqrt{3}}$
	& $-\frac{2}{3 \sqrt{3}}$
	& $\frac{1}{3 \sqrt{3}}$
	\\  
	$\Sigma^{*,0} \to \Sigma^0 \gamma$            
	& $-\frac{4}{\sqrt{3}}  F^Q$
	& $\frac{2}{\sqrt{3}}  (  D^Q - 2 F^Q)$
	& $\frac{1}{\sqrt{3}}  (D^Q - F^Q)$
	& $-\frac{1}{6\sqrt{3}}$
	& $\frac{1}{3\sqrt{3}}$
	& $-\frac{1}{6\sqrt{3}}$
	\\  
	$\Sigma^{*,0} \to \Lambda \gamma$            
	& $-\frac{4}{3}  ( 2 D^Q - 3 F^Q)$
	& $-\frac{2}{3}  ( D^Q - 6 F^Q)$
	& $\frac{1}{3}  ( D^Q + 3 F^Q)$
	& $0$
	& $0$	
	& $0$
	\\  
	$\Sigma^{*,-} \to \Sigma^- \gamma$            
	& $0$
	& $0$
	& $0$
	& $0$
	& $0$	
	& $0$
	\\  
	$\Xi^{*,0} \to \Xi^{0} \gamma$            
	& $-\frac{2}{\sqrt{3}} ( D^Q - F^Q) $
	& $-\frac{4}{\sqrt{3}}  ( D^Q - 2 F^Q)$
	& $\frac{8}{\sqrt{3}}  F^Q$
	& $\frac{1}{3 \sqrt{3}}$
	& $-\frac{2}{3 \sqrt{3}}$
	& $\frac{1}{3 \sqrt{3}}$
	\\  
	$\Xi^{*,-} \to \Xi^- \gamma$            
	& $0$
	& $0$
	& $0$
	& $0$
	& $0$	
	& $0$
\end{tabular}
\end{ruledtabular}
\end{table} 
\begin{table}
\caption{\label{t:QclebschCD}
The $SU(3)$  coefficients $C_{XX'}$ and $D_{XX'}$ in \QCPT.}
\begin{ruledtabular}
\begin{tabular}{c | c  c  c | c  c  c }
	& \multicolumn{3}{c |}{$C_{XX'}$}                         
              &   \multicolumn{3}{c}{$D_{XX'}$}    \\
        & $\eta_u \eta_u$ & $\eta_u \eta_s$&$\eta_s \eta_s$
           & $\eta_u \eta_u$ & $\eta_u \eta_s$  &$\eta_s \eta_s$     \\
	\hline
	$\D \to N \gamma$          
	& $0$
	& $0$
	& $0$
	& $0$
	& $0$	
	& $0$
	\\  
	$\Sigma^{*,+} \to \Sigma^+ \gamma$  
	& $-\frac{2}{3\sqrt{3}}$
	& $\frac{1}{3\sqrt{3}}$
	& $\frac{1}{3\sqrt{3}}$
	& $-\frac{2}{\sqrt{3}} F^Q$
	& $\frac{1}{\sqrt{3}} (D^Q+F^Q)$	
	& $-\frac{1}{\sqrt{3}} (D^Q-F^Q)$
	\\  
	$\Sigma^{*,0} \to \Sigma^0 \gamma$            
	& $0$
	& $0$
	& $0$
	& $\frac{2}{\sqrt{3}}F^Q$
	& $-\frac{1}{\sqrt{3}}(D^Q+F^Q)$	
	& $\frac{1}{\sqrt{3}}(D^Q-F^Q)$
	\\  
	$\Sigma^{*,0} \to \Lambda \gamma$            
	& $0$
	& $0$
	& $0$
	& $-\frac{4}{3} D^Q+2F^Q$
	& $\frac{5}{3} D^Q-F^Q$	
	& $- \frac{1}{3}D^Q-F^Q$
	\\  
	$\Sigma^{*,-} \to \Sigma^- \gamma$            
	& $\frac{2}{3\sqrt{3}}$
	& $-\frac{1}{3\sqrt{3}}$
	& $-\frac{1}{3\sqrt{3}}$
	& $\frac{2}{\sqrt{3}}F^Q$
	& $-\frac{1}{\sqrt{3}}(D^Q+F^Q)$	
	& $\frac{1}{\sqrt{3}}(D^Q-F^Q)$
	\\  
	$\Xi^{*,0} \to \Xi^{0} \gamma$            
	& $0$
	& $0$
	& $0$
	& $\frac{1}{\sqrt{3}}(D^Q - F^Q)$
	& $-\frac{1}{\sqrt{3}} (D^Q+F^Q)$	
	& $\frac{2}{\sqrt{3}}F^Q$
	\\  
	$\Xi^{*,-} \to \Xi^- \gamma$            
	& $\frac{1}{3\sqrt{3}}$
	& $\frac{1}{3\sqrt{3}}$
	& $-\frac{2}{3\sqrt{3}}$
	& $-\frac{1}{\sqrt{3}}(D^Q - F^Q)$
	& $\frac{1}{\sqrt{3}} (D^Q+F^Q)$	
	& $-\frac{2}{\sqrt{3}}F^Q$
\end{tabular}
\end{ruledtabular}
\end{table} 
Note that the symmetry between the
$\S^{*,+} \to \S^+ \gamma$ and 
$\Xi^{*,0} \to \Xi^0 \gamma$ transitions
as well as the
$\S^{*,-} \to \S^- \gamma$ and 
$\Xi^{*,-} \to \Xi^- \gamma$ transitions
that holds
in \CPT\ and \PQCPT\
is now broken by singlet loop contributions.

\section{\label{sec:conclusions}Conclusions}
We have calculated the 
baryon octet to decuplet transition form factors in \QCPT\ and \PQCPT\
using the the isospin limit of $SU(3)$ flavor
and have also derived the result for
the nucleon doublet in two flavor \PQCPT\
away from the isospin limit.
Extrapolating lattice calculations that employ the quenched or
partially quenched approximation can only be done
by using their respective low-energy theories,
\QCPT\ and \PQCPT.
Whereas PQQCD can be smoothly connected to QCD,
QQCD exhibits pathological behavior,
in particular QQCD observables are usually more
divergent in the chiral limit than in QCD.  This stems from the
fact that new operators not present in QCD must be included
in the QQCD Lagrangian.

For the decuplet to octet transition form factors
our NLO \QCPT\ results are not more divergent than their
\CPT\ counterparts:
$G_1,G_2\sim\a+\b\log m_Q$ and
$G_3\sim\a$.
This, however, does not mean that this result is free of 
quenching artifacts.
The quenched transition moments pick up contributions
from hairpin loops.
A particular oddity is that the quark mass dependence
of the $\Sigma^{*,-}$ and $\Xi^{*,-}$ quenched
transition moments
is solely due to the singlet parameter $\mu_0^2$;
even worse, $G_3^Q(q^2)=0$ at this order.
These transitions thus present extremes of the
quenched approximation in agreement with the
quenched lattice data of~\cite{Leinweber:1993pv} where 
the $\Sigma^{*,-}$ and $\Xi^{*,-}$ E2 moments were
found to be
significantly different from the other transitions.
In contrast to \QCPT\ results, 
our \PQCPT\ results will enable
not only the extrapolation of PQQCD lattice simulations
of the transition moments but also the extraction of predictions
for the real world: QCD.

\begin{acknowledgments}
We would like to thank Martin Savage
for very helpful discussions and for useful comments on the manuscript.
This work is supported in part by the U.S.\ Department of Energy
under Grant No.\ DE-FG03-97ER4014.
\end{acknowledgments}

\appendix*

\section{\label{s:su2} $\D\to N\g$ Transitions in $SU(2)$ flavor
         with non-degenerate quarks}
In this Appendix, we repeat the calculation of the 
transition moments for the case of $SU(2)$ 
flavor with non-degenerate quarks, i.e.,
the quark mass matrix reads
$m_Q^{SU(2)} = \diag(m_u, m_d, m_j, m_l, m_u, m_d)$.
Since defining ghost and sea quark charges is constrained only by the 
restriction that QCD be recovered
in the limit of appropriately degenerate quark masses, 
the most general form of the charge matrix is
\begin{equation}
  \cQ^{SU(2)} 
  = \diag\left(\frac{2}{3},-\frac{1}{3},q_j,q_l,q_j,q_l \right) 
.\end{equation}
The symmetry breaking pattern is assumed to be 
$SU(4|2)_L \otimes SU(4|2)_R \otimes U(1)_V
 \longrightarrow
 SU(4|2)_V \otimes U(1)_V$.
The baryon field assignments are analogous to the case of
$SU(3)$ flavor.
The nucleons are embedded as
\begin{equation} \label{eqn:SU2nucleons}
 \cB_{ijk}=\frac{1}{\sqrt{6}}\left(\e_{ij} N_k + \e_{ik} N_j\right)
,\end{equation}
where the indices $i,j$ and $k$ are restricted to $1$ or $2$ and 
the $SU(2)$ nucleon doublet is defined as
\begin{equation}
  N = \left(\begin{matrix} p \\ n \end{matrix} \right) 
\end{equation}
The decuplet field $\cT_{ijk}$, 
which is totally symmetric, 
is normalized to contain the $\Delta$-resonances 
$T_{ijk}=\cT_{ijk}$ with $i$, $j$, $k$ restricted to 1 or 2
and
$\cT_{111} = \D^{++}$.
The construction of the octet and decuplet baryons containing 
one sea or one ghost quark is analogous to the $SU(3)$ flavor
case~\cite{Beane:2002vq}
and will not be repeat here.

The free Lagrangian for $\cB$ and $\cT$ is the one in 
Eq.~(\ref{eqn:L})
(with the parameters having different numerical values than
the $SU(3)$ case).  
The connection to QCD is 
detailed in~\cite{Beane:2002vq}.
Similarly, the Lagrangian describing the interaction of the 
$\cB$ and $\cT$ with the pseudo-Goldstone bosons is 
the one in 
Eq.~(\ref{eqn:Linteract}) that can be matched to the familiar one in QCD
(by restricting the $\cB_{ijk}$ and $\cT_{ijk}$ to the $qqq$ sector),
\begin{eqnarray}
  \cL 
  &=&
  2g_A{\ol N}S^\mu A_\mu N+g_1{\ol N}S^\mu N\tr(A_\mu) 
  + 
  g_{\D N} \left( \ol{T}{}^{kji}_{\nu} A_{il}^{\nu} N_j \e_{kl} 
                      + \text{h.c}
                 \right) \nonumber \\
  &&+
  2 g_{\D \D} \ol{T}{}^\nu_{kji} S_\mu A^\mu_{il} T_{\nu,ljk} 
  + 
  2 g_X  \ol{T}{}^\nu_{kji} S_\mu  T_{\nu,ijk} \tr (A^\mu)  
,\end{eqnarray}
where one finds at tree-level 
$g_1=-2(D - F)$, $g_A = D + F$,
${\mathcal C} = - g_{\D N}$, and $\mathcal{H} = g_{\D \D}$,
with $g_X = 0$. 
The leading tree-level operators which contribute to $\D\to N\g$
have the same form as in Eq.~\eqref{eqn:Lbob},
of course the low-energy
constants have different values.

Evaluating the transition moments at NLO in the chiral expansion 
yields expressions identical
in form to those in Eqs.~\eqref{eqn:G1}--\eqref{eqn:G3} 
with the SU$(2)$ identifications made for $\mathcal{C}$,
$\mathcal{H}$, $D$, and $F$. 
For the $SU(2)$ coefficients in 
\CPT\ one finds $\b_X^B=g_A/\sqrt{3}$ and $\b_X^T=5/(3\sqrt{3})$
for the $\pi^\pm$.
The corresponding values for the case of \PQCPT\
appear in
Table~\ref{t:clebschSU2}.
\begin{table}
\caption{\label{t:clebschSU2}
The $SU(2)$ coefficients $\beta_X^B$ and $\beta_X^T$ in \PQCPT\ 
         for $\Delta \to N \gamma$.}
\begin{tabular}{r | c  c }
\hline\hline
&  $\beta_X^B$  & $\beta_X^T$ \\
\hline
   $uu$     & $\frac{1}{3 \sqrt{3}} (2 - 3 q_j)$ & $-\frac{1}{9 \sqrt{3}}(2 - 3 q_j)$ \\   
   $ud$     & $\frac{1}{\sqrt{3}}\,[1 + q_j - q_l + 2 \frac{g_A}{g_1}]\; $ & $\frac{1}{3 \sqrt{3}} (4 -  q_j + q_l)$ \\   
       $dd$     & $\frac{1}{3 \sqrt{3}} (1 + 3 q_l)$ & $-\frac{1}{9 \sqrt{3}}(1 + 3 q_l)$ \\  
       $ju$     & $-\frac{1}{3 \sqrt{3}} (2 - 3 q_j)$ & $\frac{1}{9 \sqrt{3}}(2 - 3 q_j)$ \\  
       $lu$     & $-\frac{1}{3 \sqrt{3}} (2 - 3 q_l)$ & $\frac{1}{9 \sqrt{3}}(2 - 3 q_l)$ \\
       $jd$     & $-\frac{1}{3 \sqrt{3}} (1 + 3 q_j)$ & $\frac{1}{9 \sqrt{3}}(1 +3 q_j)$ \\ 
       $ld$     & $-\frac{1}{3 \sqrt{3}} (1 + 3 q_l)$ & $\frac{1}{9 \sqrt{3}}(1 +3 q_l)$ \\  
\hline\hline
\end{tabular}
\end{table}


\end{document}